# Context-aware Authorization in Highly Dynamic Environments

**Jean-Yves TIGLI[1,*], Stéphane LAVIROTTE[1], Gaëtan REY[1], Vincent HOURDIN[1,2] and Michel RIVEILL[1]**

**[1] I3S, University of Nice – Sophia Antipolis**
**Sophia-Antipolis, France**

**[2] MobileGov**
**Sophia-Antipolis, France**

**Abstract**

Highly dynamic computing environments, like ubiquitous and pervasive computing environments, require frequent adaptation of applications. Context is a key to adapt suiting user needs. On the other hand, standard access control trusts users once they have authenticated, despite the fact that they may reach unauthorized contexts. We analyse how taking into account dynamic information like context in the authorization subsystem can improve security, and how this new access control applies to interaction patterns, like messaging or eventing. We experiment and validate our approach using context as an authorization factor for eventing in Web service for device (like UPnP or DPWS), in smart home security.[1]

## 1. Introduction

Ubiquitous computing, under the leadership of Mark Weiser's vision [1], has made computing evolve toward multi-device, multi-user, and highly dynamic environments. Miniaturization of hardware and new wireless communication networks have created new devices, worn by users or surrounding them. Due to mobility, devices appear and disappear frequently in such environments.

The major concern in ubiquitous or pervasive computing is adapting applications to users surroundings, and more generally, to their context. In these works, we focus on limiting communications between entities that are in the same context, or in a context authorized by a security system, for security purposes. Indeed, information involved in ubiquitous computing communications is often privacy-sensitive, and we need to make sure it cannot be received or intercepted by unauthorized entities.

Access control [2] relies on and coexists with authentication, authorization and audit. Authentication can be made on information or persons: it establishes who issued a piece of information, or confirms the identity of a person. However, to ensure that the identity is correct, different authentication factors shoud be used. If the person possesses the information related to each factor, it is assumed that this is the pretended person [3].

Authorization takes places both before system execution, to define policies of the security system, and after the authentication phase, to grant a principal access to the controlled system. We will study in the following section that authorization is most often static or controlled by applications, leading the users to be considered authorized for a long time. With context changes we cannot assume that a user is authorized throughout the duration of the use of an application, even if he is still authentified. We will then explore works on dynamic authorization.

## 2. Authorization

We could identify three types of authorization in existing systems: *static*, *quasi-static*, and *dynamic* authorization. Numerous works have already explored these subjects, aiming to handle more dynamicity in access control systems. In parallel, we explain how context appeared in authorization, and what properties are mandatory for dynamic environments.

### 2.1 Static Authorization

Historically, access control used static credentials to confirm user identity and was made only when entering the system. For example, the login phase of an operating system needs a login and a password to authenticate a user, and is made only when he logs in. It can also be an ID card, a fingerprint pattern, or an identification token. Infrastructure information is sometimes used to authenti-cate users. For example, the Network File System (NFS) access control uses, in its default configuration, the IP address of a client to grant him access, as long as he still uses the file system.
Classical authorization rule, also called access policy or condition, expressed that a *subject*, also called *principal* [2, 4], can access to an *operation* on a particular *object*. This can take various forms, for instance the subject can be a group of users as in the well-known role-based access control systems (RBAC), the object can be a computer, a service or an object instance.
We model the access control process with finite state diagrams. In Figure 1, a subject wants to use a system, and he has to authenticate himself in the first place. Since this is *static* authorization, if authentication is correct and matches an authorization rule, he stays authorized and considered trusted until he logs off.

[1] This work is currently supported by ANR project ANR-08-VERS-005
* Currently delegated as INRIA researcher in the team PULSAR

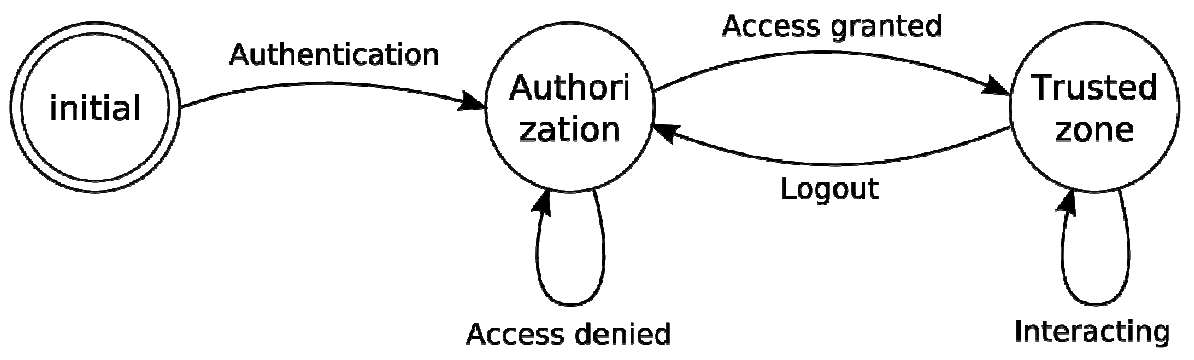

Fig. 1 Static authorization.

## 2.2 Quasi-static Authorization

More than ten years ago, static information for authentication and authorization began to be seen as a limitation in several domains. In distributed computing for example, with Nitsche *et al.* [5] and Cholewka *et al.* [6], the task being done could affect access control on some objects. The task was extracted from the workflow of the application, and this dynamic information was considered as contextual information for the application.
When context becomes a part of the authorization process, it is not only subject's identity that leads to a granted authorization anymore. Access control evolves from identity-based towards attribute-based and context-based authorization [7]. Since this kind of information is dynamic by nature, and especially in mobile or ubiquitous environments, the authorization must not be granted forever. Besides, in these systems, operations on which access control applies are generally finer-grained, which also requires to enforce authorization rules more often.
Moreover, separation of authentication and authorization process for dynamic information introduction in the access control process is put forward. In [6], users do not receive an authorization if context is not valid, even if they are authenticated. In RBAC systems, the same pattern appears [8]: a user is granted membership of a role but is not able to perform operations until infor-mation required for dynamic authorization is validated.
Later popularized by Web applications, session management has emphasized what we call *quasi-static* authorization. In these systems, credentials and other access control conditions are rarely changed compared to the lifespan of applications. Authorization is made at first access of the system, and periodically renewed to keep users authorized in case of information change in authentication or authorization information. This mechanism is called leasing, and also often used in publish/subscribe systems [9]. We modelled it in Figure 2.

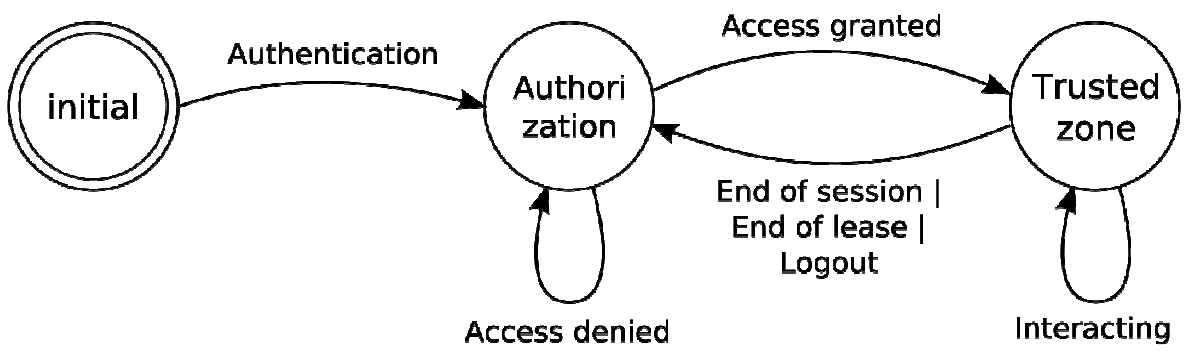

Fig. 2 Quasi-static authorization.

The main difference with *static* authorization is that a transition appears from authorized to unauthorized states. Whenever the lease expires, the subject has to be authorized again to return in trusted state. It is the application that defines for how long the subject is considered trusted.
*Quasi-static* authorization prevents users to be connected to a system forever. Changes in credentials, in the environment or the introduction of a new authentication factor in the access control system eventually lead to user's credential reevaluation and authorization policy reenforcement. As an example of such system in industry, we can cite Mobilegov Access Control [10] that uses infrastructure-based authentication in addition to password based authentication for different kind of systems.

## 2.3 Dynamic Authorization

*Static* and *quasi-static* authorization systems are inadequate for highly dynamic environments in which user's context is an important concern, and is already a part of applications. As we already seen with *quasi-static* authorization, not using contextual information in security concerns could lead to granting a user access without considering his condition [11]. Besides, handling context in access control systems has been identified as a key concern of pervasive computing evolution [12]. More precisely, it is the authorization that should be related to context-sensitivity [13].
Contextual information is highly dynamic, because the user is likely to be moving, as much as other users in the same ambient space, with their attached devices. Yet, sensors can be fixed in the physical infrastructure, like temperature or light sensors. Contextual dynamic information is used to invalidate subject's authorization, even if he is still identified by standard authentication factors.

Thus, we introduce the *dynamic* authorization model for environments in which it is needed to frequently check if subjects are authorized due to changes in dynamic information used for authorization. This open gates to considering highly dynamic contextual information to be used in the access control process. As opposition to *static* and *quasi-static* authorization, *dynamic* authorization requires to be rechecked according to changes in dynamic information. It is necessary to dynamically modify access permissions granted to subjects when context information or when software infrastructure change.

While in *static* and *quasi-static* authorization subjects were trusted as long as they were logged or for a predefined time, in dynamic authorization, authorization must be checked at each operation in the system. This can be done in two ways:

- The first would be to reduce the lease time near zero, and thus needing subjects to authenticate and subscribe all the time. Lease time has to be adapted to system's reactivity, which is around one second for ubiquitous computing applications for example. This is very inefficient and consequently a bad solution for embedded devices populating ubiquitous computing environments,
- The second, to be more efficient, would need the system be notified about subject's context all along his use of the system. In that case, the system could react on context changes by enforcing authorization policies to determine if the subject is still authorized and can be kept or not the trusted area. We modelled this system in Figure 3.

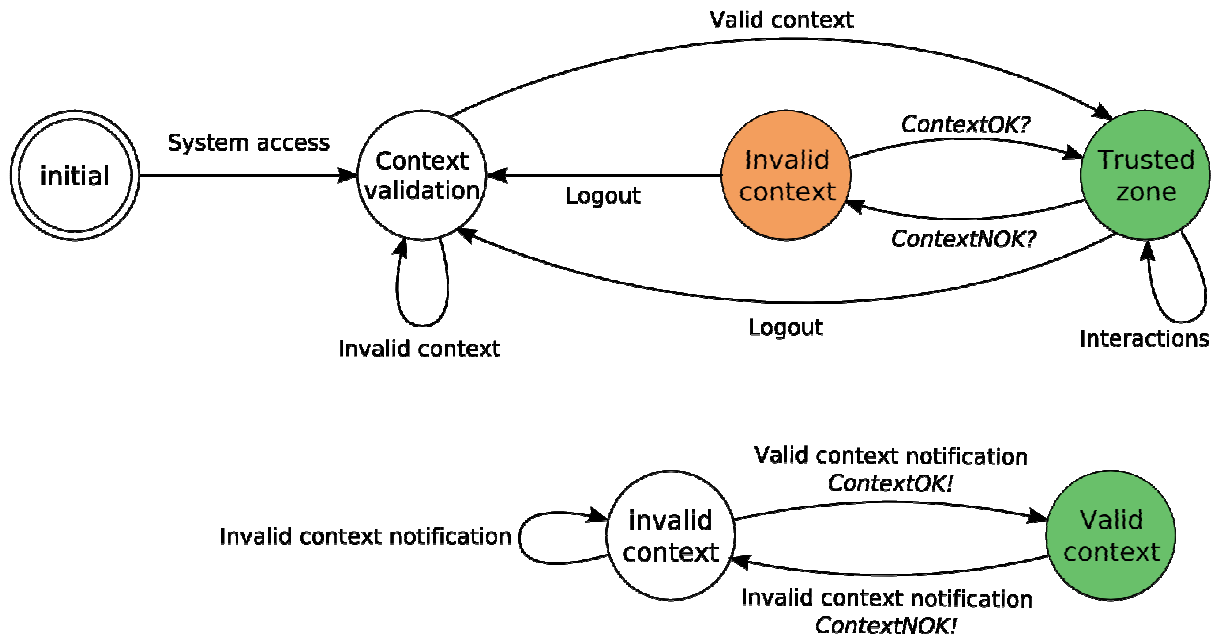

Fig. 3 Event-driven dynamic authorization.

With this second solution, trusted zone exit and re-entery are context-driven. Since the dynamics of the context and of the application are different, the access control process is highly reactive. *Quasi-static* and *static* authorization process, in contrast, were driven by the application. It also means that a new concern appears as a part of the access control process: dynamic information management.

New issues appear with management of contexual information for authorization purpose. For example, the Context-aware provisional-based access control (CAPBAC) [7], formalizes how contextual information should be managed in an access control system. In a such system, actions can be performed between the authentication validation and granting the access to the subject. Information has to be present in the system when context-based authorization rules are evaluated.

Cuppens *et al.* [15] also formalized a context-sensitive authorization process using logic rules. They defined how contextual constraints of five types could be added to the system: temporal, spatial, prerequisite depending on subject's action's history, user-declared goals and provisions. However, the way all information is gathered is predefined and they don't address how the system can adapt to changes in the infrastructure, like appearance of new information sources. McDaniel, in other formalization works, states that policy conditions are external complex general purposes programs [13] and thus can adapt to infrastructure changes.

More than purely contextual issues, like the type of contextual information and the way it is collected, security issues are also linked with these new challenges:

- *How to ensure that dynamic information is authentic?* As stated Kindberg and Zhang, in their experience in the location-aware mobile computing CoolTown project [3]: when using contextual information for access control, the authentication of the data itself must be done. Indeed, dynamic data are provided by sensors, and they can be simulated of falsified if protocols are not constrained as in [3]. In some cases with group behaviors, information can also be correlated with surrounding entities' to check forged information [4]. If sensors are not able to sign information, it has to be authenticated when users collect it. A trusted observer has to collect the same information than users in order to authenticate it, and verify that it is this information that is used by users to access the system.
- *What about privacy?* Of course, placing a trusted entity in users computing environment can be recusant. Westin [16] defined privacy as "*the ability to determine for ourselves when, how, and to what extent information about us is communicated to others*". If the trusted entity describes precisely how contextual information is used, it should be accepted by users. Furthermore, one must consider that machine-to-machine communications play a more and more important role, and that privacy in those cases is not relevant.

Table 1: Classification of context-aware access control projects

| Project | AuthZ | Information from | AuthN | Targetted system | Oper |
|---|---|---|---|---|---|
| Nitsche [5] | *N\A* | workflow (self) | 0 | medical information | = |
| Cholewka [6] | quasi | workflow (self) | 0 | task in a workflow | = |
| CS-RBAC [8] | quasi | *unspecified* | 0 | operation/method/LDAP | CL |
| Cooltown [3] | quasi | constrained channels | 1 | *irrelevant: information gathering only* | |
| OASIS [14] | dynamic | local database | 1 | general purpose | = AND |
| Antigone[13] | quasi | trusted sources | 1 | general purpose | = |
| OrBAC [15] | quasi | predefined types and entities | 0 | medical information | range |
| CA-RBAC [11] | dynamic | sensors (concept only) | 0 | resources | = |
| CSAC [4] | quasi | trusted encryption-aware observers, surroundings, history | 1 | general purpose, services | = zone |
| CA-PBAC [7] | quasi | context engine (traffic, time...) | 0 | general operation (applications) | = |

- *Granted access implies information leak?* Hengartner and Steenkiste [17] state that being granted access to a system using context-sensitive authorization can leak some private information. Since access is limited to a predefined context, observing that a user obtained access is equivalent to know that he is in that context. Their solution is to introduce secret fields in contextual conditions. [18] use confidentiality in authorization rules. Our idea is to define several applicable conditions, possibly subject-dependant.

A good example of dynamic authorization based system is the work of Bacon *et al.*, who introduce in [14] the OASIS (Open Architecture for Securely Interworking Services) Role-Based Access Control. It uses credentials that a user possesses, along with *side conditions* that depend on the state of the environment, to authorize him to activate a number of roles. In their model, predicates can be used for environmental constraints or context-sensitive information. Environmental constraints can be checked by any entity in the environment of the application, thus it can authenticate dynamic information used for authorization.

### 2.4 Synthesis

The Table 1 summarizes the types of authorization and contextual information featured by the projects taking into accound context in an access control process. The *AuthZ* column refers to the dynamicity of the authorization as we studied it; the *Information from* column summarizes how contextual information is obtained and if it is authenticated in the *AuthN* column ("1" if it is). The granularity of the access control, or the domain of application is found in the *Targetted system* column. Finally, the *Oper* column describes what operators can be used to define rules based on contextual information: "=" denotes the equality only, "CL" means comparizon and logic operators and thus more complex rules, "= AND" for equality of information and composition with logic AND rules, "range" means that ranges of values are allowable, "= zone" is equality on coordinates, so we guess there is a range mechanism.

Most projects marqued as *quasi-static* are not conceived for highly dynamic environments. Contextual information is not triggering changes in authorization granted or rejected to users. Applications have to poll for changes when they decide it is needed. We can also notice that only a few projects are concerned about contextual infor-mation authentication. As rightly stated by McDaniel [13], condition security has been largely forgotten by recent works in authorization policies. And beyond that, either the source of information is restricted to some predefined types and entities, either there is a confiden-tiality leak in the system.

We see from the table that no project is able to handle all required concerns for a general purpose dynamic authorization system. It is a complicated task to manage contextual information sources, gather contextual information, verify information authenticity, define dynamic authorization rules, adapt security system to available sources, and actually authorize subjects to access the system in the same software architecture or framework.

We will study our proposition tackling these challenges in following sections, but before that, in the next section, we will focus on the nature of operations for which authorization rules are enforced. In highly dynamic environments, messages are often the most basic operation that has to be controlled.

## 3. Access Control in Interaction Patterns

The granularity of operations access control systems aim to limit to subjects can vary a lot. Some systems control access to a physical place, some other to networks, or to a machine. *Dynamic* authorization aims finer granularity operations, like messages. Messages in highly dynamic environments are often private, and an application of *dynamic* authorization is to manage confidentiality over a messaging channel.

In a more general way, dynamic authorization can be applied to all applications in which context can be used to limit something. For example, one could say: if you are not in my context, you won't receive my information; if you are not in my context, you won't even know who I am, and what I can provide to you. Everything in distri-buted computing is about interactions between entities, and using access control on messages allows us to create all applications based on context-aware access control. More than security, context-sensitive messaging has implications on power consumption of mobile devices that populate highly dynamic environments.

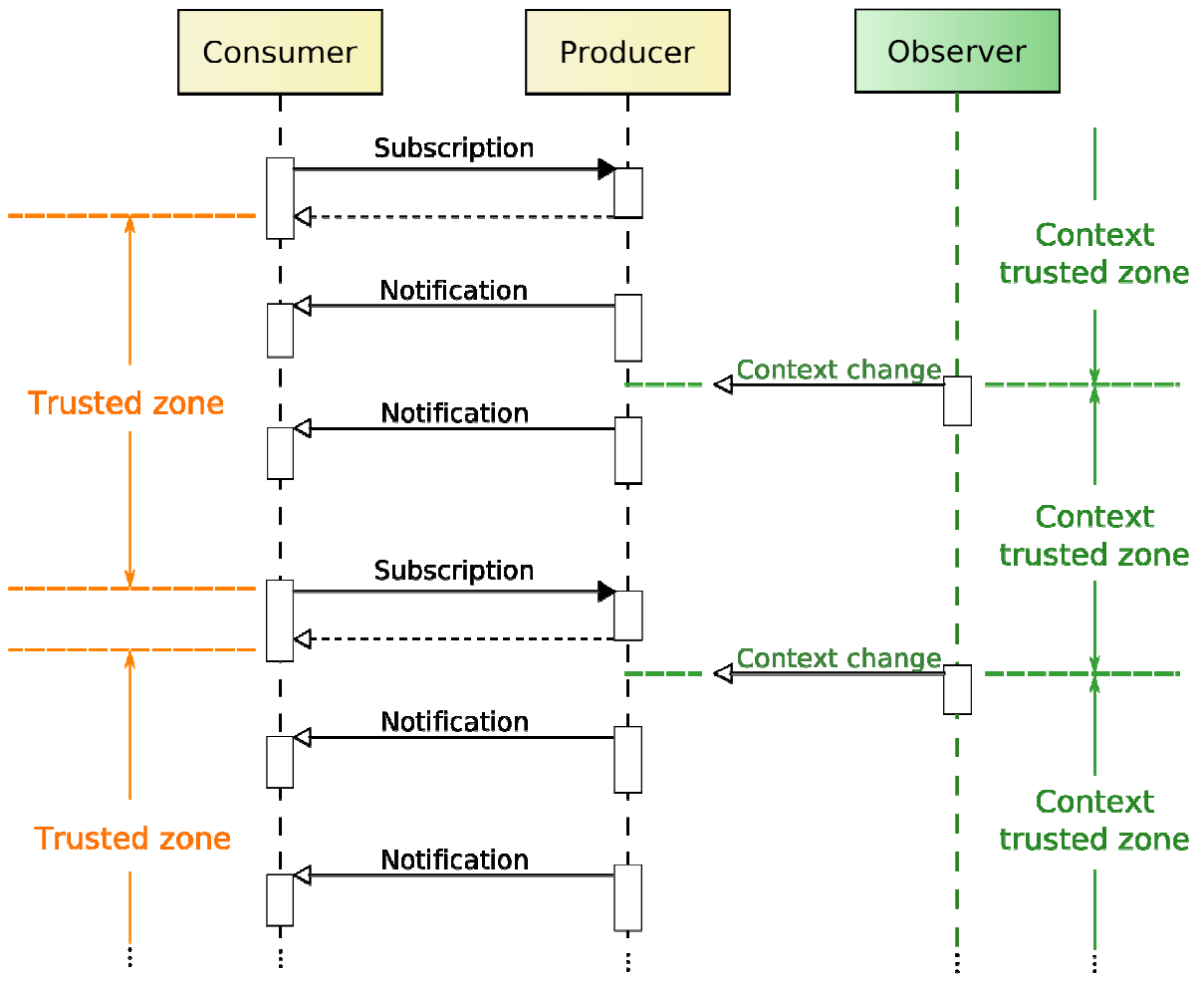

Fig. 4 Sequence diagram of trusted zones in a publish/subscribe pattern.

To emphasize where the problem is, we explain it for the well known publish/subscribe pattern [9] (Figure 4). Publish/subscribe systems are based on two kinds of inter-actions: the subscription and notifications. Notifications allow the event producer to send information to subscribed entities that he does not necessarily know beforehand.
The subscription is a synchronous process, like a request-response pattern. It is used by consumers to register their interest to a specific event channel and to give informa-tion about the connection that will be used to send events.
Notification is a purely asynchronous process, made of messages sent by the producer to the consumer. This process thus needs the consumer to be authorized to receive events. Since access control requires the consumer to send authentication and authorization information to the producer, it is practically done when the client subscribes.
However, since following interactions are only one way messages, authorization of the subscriber cannot be verified. For *static* authorization, as we have seen, this is not a problem because after subscription, it is not supposed to have changed or it is not important for system's security. With *quasi-static* authorization, the subscription is accepted only for a defined validity time: the lease. Subscriber is trusted only for this time, and has to renew his subscription and access, before the end of the lease, to avoid a service interruption. We call this lease of trust the trusted zone (Figure 4). A modelization of this process is equivalent to the *quasi-static* authorization diagram in Figure 2, in which the `access granted` transition is replaced by subscription and access control.
*What can be done for dynamic authorization of the recipient?* Figure 4 helps to understand where the problem exactly is. The context observer notifies when context validity has changed. The observer lifeline is displayed only for comparizon purposes with the standard publish/subscribe leasing mechanism. Context changes happen while the consumer is in the trusted zone. With *dynamic* authorization, the producer would reenforce the authorization conditions as soon as an event from the context is received. **With *quasi-static* authorization, the consumer is still able to receive notifications, even while his context is not authorized.**
We define the *context trusted zone* as the period during which the producer can be certain that the consumer is authorized by its context. Contrary to the trusted zone of usual interaction patterns in which information leak can occur, the *context trusted zone* ensures *confidentiality of messages*.
Bacon *et al.* [19] already explored access control based on contextual information in publish/subscribe systems; with more details, they focus on a Message Oriented Middleware (MOM) for large scale architectures with multiple administration domains. They use a dedicated security infrastructure for credential management (OASIS RBAC [14]). They apply access control only on event brokers since they are the link to inter-domain networks. Their solution is thus based on managing security through a layer below the application layer: the transport layer.
In the next section, we describe our contribution, how we handle *dynamic* access control for asynchronous communications recipients, in the application layer, and without needing a specific infrastructure for security or message management purposes.

## 4. Context-based Dynamic Authorization

We have seen in section 2 that in highly dynamic environments, *static* or *quasi-static* authorization should not be used for two reasons: some contexts are incom-patible with the authorization granted in first place, and context evolves with a different dynamic than the application. We also have seen that an efficient solution would require a trusted entity from the security system to be placed in users' context to ensure the authentication of dynamic information used for access control.
We don't want to focus only on a specific problem of the access control process, as most works we have studied did. Indeed, a context-sensitive access control architecture should deal with all issues related to access, collection, storage, processing and distribution of context information [4].
Therefore, we begin by presenting our context model, and how we manage contextual information (4.1). Then we present our dynamic authorization model with context-sensitive rules (4.2) and we explain how it applies to all kinds of interaction patterns (4.3).

## 4.1 Context Model

As a basic context-aware system, contextual information collection is done using context *observers* [20]. We don't want to delegate context management to a context toolkit or a context server, because that would prevent us from authenticating contextual information. Using such external information inside a security system like access control requires the establishment of a chain of trust all along the processing of information.
Context observers are entities of the infrastructure that provide contextual information. We assume that, as most entities in higly dynamic environments, they can notify new information by sending events. They may not be sensors, but rather standard devices. We do not limit the type of information or the list of context observers. It can be *any kind* of environmental information, like temperature, localization, time, infrastructure information like what devices are present on the network, or any other information that can be authenticated, like history of user actions.
Observers are dynamic entities. They appear and disappear from the software infrastructure like any other device or application. Consequently, more than taking into account dynamic contextual information, we need to adapt to the high dynamicity of relevant observers presence. Indeed, before designing complex context management and context-aware applications for dynamic environ-ments, we need to be able to find observers that will provide contextual information. Thus, observers are dyna-mically discovered. Furthermore, context is built from distributed and decentralized information.
An implication of such infrastructure dynamicity is that we will be able to define what authorization conditions should be enforced, depending on what observers are present. We define contextual conditions for authorization as a tuple of `<context, operation, object>`. The defined context has to be valid in order to authorize access to the operation of the object.
Every device that will be used as an observer provides raw information. This information has to be processed in order to know if it corresponds to a valid context. We define φ functions as partial conditions, that validate a simple contextual information by returning a boolean value of validity. They are similar to McDaniel's conditions, which are trusted external programs at the charge of the programmer [13]. Thus, each observer $Ob_i$ is attached to a $\varphi_i$ function. The set Φ of available $\varphi_i$ for authorization rules dynamically depends on the presence of $Ob_i$ in the infrastructure.
Finally, authentication of contextual information is made by placing a trusted party in the context of the entity to verify that the information that is sent to the φ functions and the authorization condition enforcement is authentic. Placing it in its context provides access to the same information than it obtains. A big advantage of this authentication technique is that any device can become a contextual observer, and not just predefined or constrained ones.

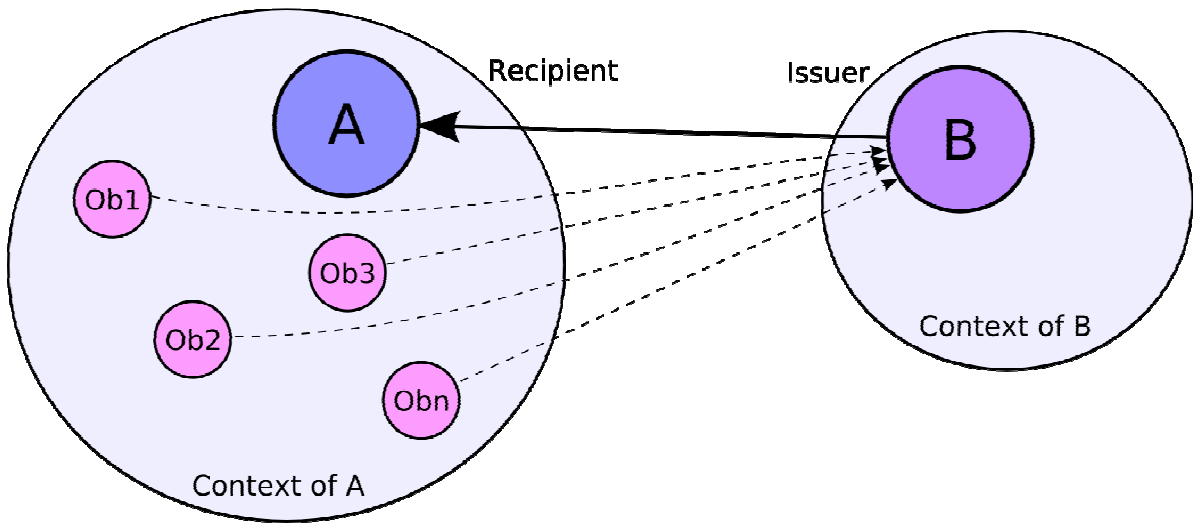

Fig. 5 Asynchronous communication and contexts.

## 4.2 Dynamic Authorization Model

Since the basic operation in our approach is the message, we represent in Figure 5 a simple application example of message sending between entities B and A. Rounds tagged with $Ob_i$ represent context observers in A's context. The problem is described as follows: when B sends a one-way message to A, how can it ensure that A is in a context in agreement with B's policy for recipients?
When observers are present, authenticated, and that the value of the contextual information they provide corresponds to an authorized value, the access is granted. The enforcement of an authorization condition is kept simple, based on a pre-requisite Π and the actual enforcement operation Γ, here expressed as logic rules:

$$\Pi \equiv \text{valid}(Ob1) \wedge \ldots \wedge \text{valid}(Obn)$$

$$\Gamma \equiv \varphi_1 \wedge \ldots \wedge \varphi_n$$

These rules imply that when an observer is missing, or is not able of being authenticated, the enforcement Γ won't be done. Conditions based on φ are removed from the list of active conditions when $Ob_i$ is missing. When all observers can be properly used in the security process, the authorization is only granted when all φ functions which are a part of this condition return valid context information. Since they provide a boolean information, a simple AND logic operator is used for evaluation.

Conditions are written as part of the authorization process to grant access to subjects. Several rules should exist for one subject, each based on different observers. This allow to grant subjects access based on contextual information while they evolve in not already discovered environments. It also means that when several conditions are applicable (when their Π is true), they should not specialize the contextual conditions, but offer alternatives, in order to avoid authorization revocation if an observer of one condition becomes out of valid context. Consequently, between each condition, a `OR` logic operator is used for evaluation.
To summarize how conditions are written, if several observers need to be used to define one context, their information, through the use of their φ function, must be used in a single condition. On the contrary, if several observers define several contexts, and that mere subsets of these observers can define a valid context for authorization, then several conditions should be written. Figure 6 represents on one side observers and their associated φ function of evaluation, and on the other side contextual authorization conditions based on φ functions.

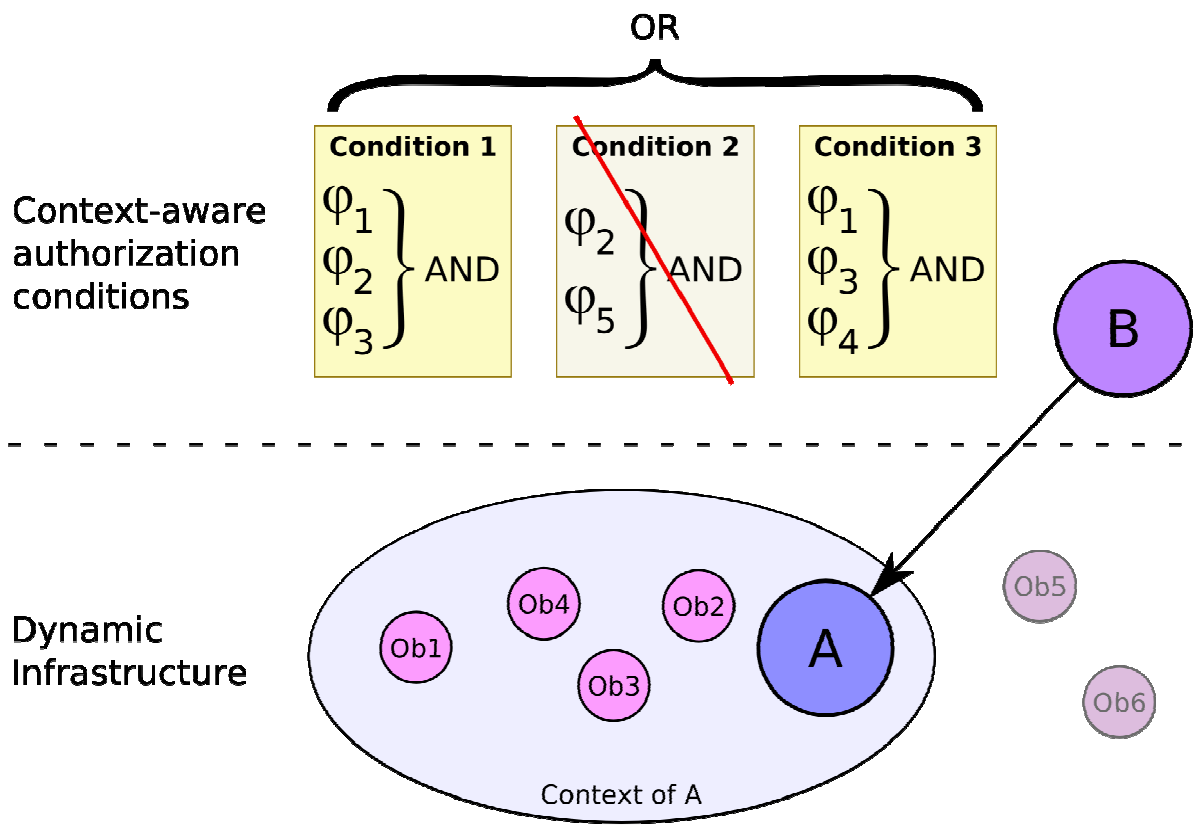

Fig. 6 Authorization conditions and based on observers availability.

Once a subject is authenticated, the status of its authorization is bound to the status of validity of information provided by observers that are in activated conditions in the system. Thanks to event commu-nications between observers, φ functions and conditions, authorization are granted and revoked dynamically and reactively, which is the very essence of *dynamic* authorization.
*Our contribution is to dynamically add trusted context observers in the context of entities, that notify the controlling entity from changes in contextual information that are used for end-to-end access control.*

## 4.3 Application to All Interaction Patterns

We already took the example of publish/subscribe systems to describe how the problem could appear. However, other interaction patterns may have the same information leak issue on context changes.

**Synchronous interactions:** The most representative synchronous interaction pattern is the request/response mode, used in method invocation and Remote Method Invocation (RMI). In this pattern, two messages are used for each interaction. The first is sent by the consumer to request the execution of some procedure on the producer, possibly with parameters. The second message is sent by the producer to the consumer with the result of the processing.
We depicted in Figure 7 a *dynamic* authorization example for request/response patterns. As a synchronous pattern, it is usually supposed more secure than asynchronous patterns. Subjects are often considered trusted during a method invocation, for example in [5]. But as we see in the figure, the same problem appears in this pattern too.
The first message is used by the consumer to send his contextual or authentication information in order to grant access to the method invocation. A context change can occur after this message has been sent, placing the consumer in an unauthorized context. Moreover, the execution time of the method may take several seconds, or even minutes. In mobile environments, the infrastructure changes often, and these circumstances can happen quite frequently.
With *dynamic* authorization, as soon as the context of the consumer gets unauthorized, the procedure processing can be stopped to spare resources, and the consumer is sent an access denied message. In contrast, with quasi-static authorization, the producer would not notice that the context has changed, and he would consider the consumer to be still in a trusted zone. The message potentially containing confidential information would be leaked.

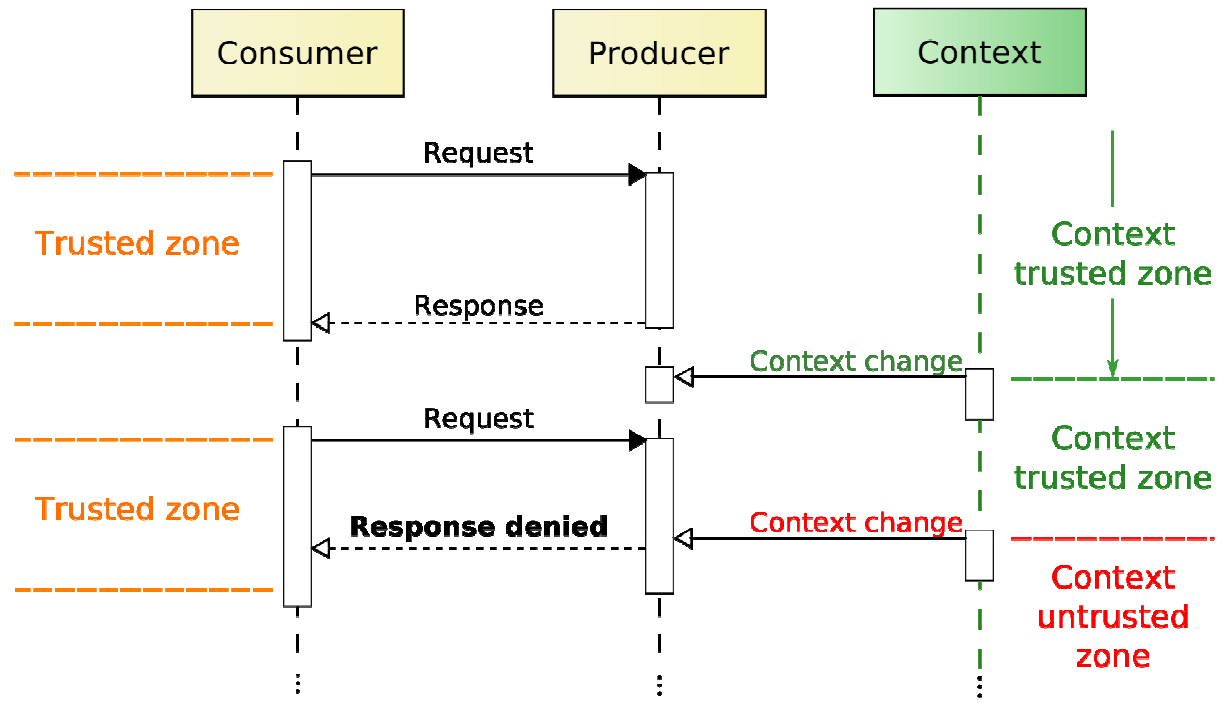

Fig. 7 Sequence diagram with context trusted zones for request/response pattern.

**Signaling and broadcasting interactions:** The third main class of interaction pattern we could identify is the signaling or broadcasting. This pattern is probably the most complicated in which access control can be handled. In DPWS (*Device Profile for Web Services*) for example,

WSDiscovery, which uses multicast messages for reactive discovery of Web services, is the only interaction scheme of DPWS that does not handle confidentiality [21]. The reason lies in the decoupling that it provides. Indeed, Eugster [9] has identified three types of coupling:

- time: the consumer and the producer have to be online at the same time. The message is not bufferized, except at operating system level if this is a distributed interaction, and is thus processed immediately.
- space: the consumer is known by the producer. In broadcasting and eventing patterns, producers and consumers are often called loosely coupled because they are not bound at design-time, nor designed specifically to execute one with each other. The space decoupling often leads to the fact that several consumers receive the producer's messages. Likewise, in complex publish/subscribe systems, there can be several producers sending messages in the same application.
- synchronization: the consumer is blocked until the producer sends the resulting message. This is typically how request/response is coupled. Asynchronous request/response actually decouples the synchronization of entities: the consumer is able to continue to execute and will be notified when the request's response is ready.

Signaling and broadcasting are decoupled in space and synchronization. Most eventing systems also have at least these two decoupling. The problem actually appears on a lower level: the transport layer. Publish/subscribe systems are space decoupled from the producer's point of view, but not from the messaging system's point of view. Indeed, consumers have to subscribe, and consequently they are known from the subscription system. *Notifications are then sent using unicast messages to consumers.*

With broadcasting, consumers cannot be known. The pattern is purely one-way, like in TV broadcasting. They are considered in a trusted zone permanently. This is exactly the same problem that appears at the application layer of a publish/subscribe system. The producer may not be aware of subscriptions, and thus cannot deal with access control for each client. If we want to handle access control at the application layer, space decoupling has to use cryptography as a means of access control.

In Bacon works [19], group cryptography is used to ensure confidentiality of events between trusted brokers. Keys are updated when principals are declared unauth-orized, and not when they unsubscribe, which makes updates happen less frequently in this kind of environ-ment. We will use the same technique to ensure that unauthorized entities cannot receive messages.

The *dynamic* authorization can be applied on any message as long as there is at least one synchronous exchange for trust establishment. For signaling, a solution still exists when the consumer is able to reach the producer: the twostep signaling. A first message is broadcasted, containing no confidential information and only a basic description of how to reach the producer. The second step is initiated by consumers registering their interest for the information, like a subscription in publish/subscribe systems. Then, for notifications (broadcasts or signals), a group key encryption is used. Only consumers in authorized context will have access to the decryption key. As soon as the context of a consumer becomes unauthorized, the group key is changed and spread to other authorized consumers.

*The dynamic authorization in eventing and in broadcasting patterns can be handled the same way because of the space decoupling they both offer. This decoupling allows us to consider these two patterns as a single problem for context-awareness and access control.* The application of this contribution to a specific infrastructure will allow us to verify it.

## 5. Application for Eventing in Web Service for Device

We chose to implement our context-sensitive authori-zation with two specific architectures and paradigms: Web service for device for the software infrastructure, and publish/subscribe systems for asynchronous communica-tion. Reasons of these choices revolve around two con-cepts: ubiquitous computing and space decoupling.

For many years, service oriented architectures (SOA) have been used in home automation, mobile, pervasive and ubiquitous computing to represent as services the sets of functionalities offered by devices. They offer lots of features discussed in [22] such as encapsulation, dynami-city, discoverability and interoperability. They evolved from standard SOA to SO A for device (SOAD) by adding two main features: decentralized reactive discovery and asynchronous communications.

Decentralized reactive discovery has been popularized by projects such as SLP (the Service Location Protocol) [23] or Jini [24]. They suppress the need of a service registry tracking all services active in a network domain. They use multicasted or broadcasted messages to notify that services appear or disappear. Asynchronous communica-tions used by SOAD like Jini are events in a publish/ subscribe scheme.

These evolutions allow to create reactive dynamic distributed applications, suitable for ubiquitous computing environments. In addition, when Web technologies are used to implement SOAD, interoperability between all entities is enabled, whether they are heterogeneous devices or simple software services. Only two implemen-tations of Web services for devices currently exist: UPnP [2] and DPWS [21]. UPnP has been created by the UPnP Forum, under the leadership of Microsoft in 1999. It has never be standardized, but is used in many objects of everyday life, like home gateways, or media centers. DPWS appeared in 2004, as a replacement for UPnP, and as a technology based on several Web services standards, like WS-Discovery or WSEventing. Its main improvement over UPnP is security, in terms of authentication and confidentiality, through the use of WS-Security.
Publish/subscribe systems use $1 \rightarrow N$ communication scheme: a publisher is able to accept several subscriptions from different clients. Thus, all consumers are notified when issuing an event. This feature will require that observers are managed for each subscriber to the eventing channel, and not for each eventing channel.

## 5.1 Service for Device Composition

To create applications from this infrastructure of services for devices, we use the Service Lightweight Component Architecture (SLCA) [25]. It allows to dynamically orchestrate and compose services for devices using light-weight components. Components are called lightweight because they execute in the same memory addres-sing space, the same process, and the same component container. The container provides the least possible technical services, also known as non-functional concerns helpers. Distribution thus has to be explicit: if a component needs to communicate remotely, it has to embed the code to do so. Obviously, we created some external tools that can generate predefined components. From Web services for devices description interfaces for example, we generate client components, that we call *proxy components*.
Containers manage assemblies of components fully dynamically. Component types can be loaded and unloaded, component instances and bindings between them can be added or removed at run-time. Proxy compo-nents are generated, loaded and instantiated dynamically and automatically. Thus, we can follow the presence of a service in a container, by adding or removing proxy components when the service appears or disappears.
Applications or new functionalities can be created from existing services on the infrastructure by managing an assembly of components inside a container. Proxy components are combined together or with purely functional components to transform information. SLCA components and services for devices communicate mostly using event-based communication patterns, which, more than decoupling entities and increasing dynamicity, will allow to react to context changes efficiently.

[2] Universal Plug and Play Forum: http://www.upnp.org/

Finally, containers can export functionalities created by component assemblies as a new web service for device using *probe components*. Each container has a dynamic functional service interface. When a probe component is instantiated or destroyed, the interface is dynamically modified: a method or an event is added or removed. Consequently, interfaces of existing services can be cloned using adequate probe components. Such services can be secured by adding functional or proxy components to the assembly. Hierarchy in the model is possible but has to use the service layer, which, moreover, allows it to be distributed.

## 5.2 Composite Service for Device Adaptation

Since compositions are based on lightweight components, service compositions are fully dynamic. A paradigm called Aspect of Assembly [22] allows to adapt composite services according to specified rules. Aspects of assembly are pieces of information describing how an assembly of components will be structurally modified, keeping blackbox property of components. Modifications include adding components and bindings between them. Aspects of Assembly consist of two parts, like regular aspects found in Aspect-Oriented Programming (AOP) [26]: pointcut and advice. Pointcuts describe to which components the modifications described by advices have to be weaved (applied).
If some of the required components expressed in a pointcut are not available, the advice won't be weaved until they become all available. Since service discovery is a reactive process and that containers notifications are events too, aspects can be weaved in response to the appearance of a service (and thus a device) on the infra-structure.
Moreover, aspects of assembly provide associativity, commutativity and idempotence properties when several aspects are enabled to be weaved at the same time [22]. This allows us to manage several aspects of assembly for different crosscutting concerns, and ensure that their composition is deterministic and that the application handles all concerns in a right way.

## 5.3 Implementation

The service for device infrastructure and SLCA are used for all parts of the application: publisher, subscriber and observers. Observers are trusted entities from the publisher's point of view thanks to dynamic insertion of authentication components with aspects of assemblies.
We created a simple example of application, modelled in Figure 8. An event publisher service, which can be a sensor or any device, is secured by the composite service on the left. The client of this secured service is a composite service to simplify the figure. This can of course be applied to already existing service clients by only modifying the location (URL) of the service used with a security proxy composite service. Observers are managed in the context of the client by another composite service, to simplify communications.

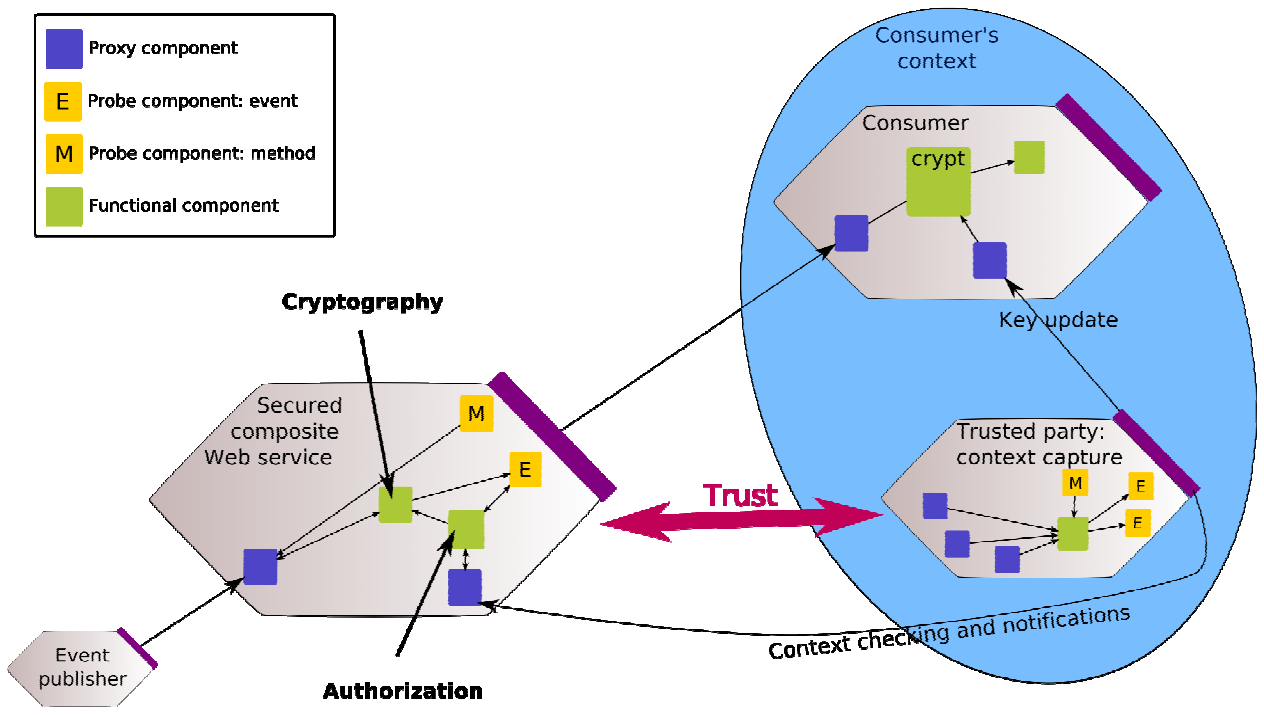

Fig. 8 Implementation using SLCA.

An idea behind the use of lightweight components in composite services is to enable adapting non-functional concerns in the same layer and the same way than the functional core of the application. We use aspects of assembly in the publisher's and subscriber's composite services to add the access control logic.

Since we manage all concerns of the application on the same layer, we cannot deal directly with subscriptions handled by the underlying service infrastructure. We have to manage authorizations for all subscribers at the application layer, as we have studied in (4.3).

Events are encrypted with a group key. When observers notify changes of contextual information, if an authorization rule becomes invalid, the security system in the composite service of the event producer changes the group key. Modifications of the key are spread to the subscribers of the event channel using the observers. Indeed, since they are in subscribers' context and they are trusted parties, they can safely deliver the new key.

Aspects of assembly are used to implement contextual conditions for authorization. They allow us to manage different conditions based on appearing and disappearing trusted observers in the environment. Thanks to properties of aspects of assembly, and as we created a new composition operator with logic OR semantics, we can enable several rules to be used at the same time for *dynamic* authorization. Even if they are enabled, they won't apply until all observers needed by the rule, defined in pointcuts, are present. The reactive discovery process of Web service for device makes adaptation of authorization rules reactive. This is useful in cases of context overlappings and transitions, or simply to ensure that access won't be denied because of slight changes in the highly dynamic infrastructure of ubiquitous computing.

## 5.4 Application Examples

As a common service middleware, our SLCA/AA implementation allows to create applications in many domains. In smart home security for example, we can create contextsensitive secured applications. According to the british *Royal Society for the Prevention of Accidents* (RoSPA), in 2002, more than 41,000 children were injured by thermal effect. In smart homes, with context-aware securization, accidents caused by children been left alone with dangerous household equipment would be reduced.

We identified several contextual information that can be used in smart home environments: time ranges, localization partitioned in rooms, presence of persons nearby others, status of persons between each other like children needing attention and adults who monitor them, and all other information provided by devices of the home, like the opening state of doors, lights on or off, temperatures, and so on. Contextual conditions can be written to take into accound this information in authorization.

Another application in the smart home domain is to allow access to the home network and its devices only to persons inside a secure place like the house, the cars of the inhabitants, or the garden, only if the front door is unlocked. All three conditions can be placed in the system, but only those relevant with observers would be activated. That would prevent intruders to crack into wireless networks as it is often the case today, since either they are not in a valid context, either don't have observers related to an authorization condition. They wouldn't either be able to falsify these observers and forge their information they provide because the trusted entity uses strong cryptography to sign information.

# 6. Validation

More than the implementation of the contribution that proves that it can rely on existing service architectures and dynamic composition, we validate our approach by calculating the reactivity of the *dynamic* authorization process; we also compare the number of messages exchanged between subjects and controller process for *quasi-static* and *dynamic* authorizations.

The process of taking into account changes in contextual information in the authorization involves several operations. Hence, the time elapsed between the variation of a contextual information and the modification of the authorization is decomposed as follows: data processing or sensing by the observer ($o$), communication between the observer and the proxy component of the event provider ($c$), processing of the $\varphi$ function ($f$), and enforcement of the condition leading to a key change in the composite authorization service ($p$). *reaction time* $= o + c + f + p$. $o$ and $p$ are local data processing and take typically less than 1 ms to execute. *f* is also executed quickly, except if it uses the history of values to filter inconsistencies. It would then need the time between two informations from the observer to propagate the new information. *c* depends on how many hops there are between the subscriber and the event provider. In ubiquitous computing, wireless networks are often used, so c may suffer from an important variance. If there is no communication problem and that φ is not buffering the information, the reaction time can be lower than 10ms. Unfortunately, in such dynamic environments, it can reach a few seconds.

In *quasi-static* authorization, like lease-based systems, the value of the lease is several orders higher. The UPnP specification, for example, recommends it to be at least half an hour. In security aware systems though, it shouldn't be less than one minute to be efficient enough and not power greedy. The reaction time would then be at maximum the value of the lease, since the authorization process is reprocessed at the same time.

The number of messages used for the authorization process in *quasi-static* authorization is periodically increased. Indeed, the leased subscription makes those messages to be send at every lease. Thus, this number follows a linear law, function of the time spent using the system. In *dynamic* authorization, messages are sent only when dynamic information is modified. It can be higher than the linear law if context validity changes more often than the lease time. Else, it can be lower in number of message sent, but still more reactive.

## 7. Conclusion and Trends

We have described a solution that allows *dynamic* authorization policies based on dynamic information to be used to manage message access control. While other systems generally use access control at a higher level, like accessing the entire system or a connection, we argue that in highly dynamic environments access control has to be finer-grained. Moreover, end-to-end security is achieved, since our contribution does not rely on a dedicated security infrastructure and is integrated in the application layer. Reactive management of dynamic information changes, like context, makes the solution efficient. Finally, context can be actually used as an improvement for access control systems in the authorization process.

Future works will study how can several observers' information can be correlated to a single φ function, in order to create more complex context-sensitive rules. We will also explore how we can generate contextual conditions at runtime, to adapt to new environments or make the access control process more transparent, as imagined in [4]. This may also be used as described in [11], to not only change users access privileges, but also adjust access permissions of resources.

**Acknowledgments**

This work is supported by the French ANR Research Program VERSO in the project *ANR-08-VERS-005* called CONTINUUM.

## References

[1] M. Weiser. "The computer for the twenty-first century". **Scientific American**, Vol. 265 No. 3, 1991, pp. 94–104.

[2] R. Sandhu and P. Samarati, "Access control: principle and practice," **IEEE Communications Magazine**, Vol. 32, No. 9, 1994, pp. 40–48.

[3] T. Kindberg, K. Zhang, and N. Shankar, "Context authentication using constrained channels," in **Fourth IEEEWorkshop on Mobile Computing Systems and Applications**, 2002, pp. 14–21, IEEE Computer Society.

[4] R. Hulsebosch, A. Salden, M. Bargh, P. Ebben, and J. Reitsma, "Context sensitive access control," in **tenth ACM symposium on Access control models and technologies**, 2005, pp. 111–119, ACM.

[5] U. Nitsche, R. Holbein, O. Morger, and S. Teufel, "Realization of a context-dependent access control mechanism on a commercial platform," in **14th International Information Security Conference (IFIP/SEC)**, 1998.

[6] D. G. Cholewka, R. A. Botha, and J. H. P. Eloff, "A contextsensitiveaccess control model and prototype implementation," in **IFIP TC11 Fifteenth Annual Working Conference on Information Security for Global Information Infrastructures**, 2000, pp. 341–350, Kluwer Academic Publishers.

[7] A. Masoumzadeh, M. Amini, and R. Jalili, "Context-aware provisional access control," **Lecture Notes in Computer Science**, Vol. 4332, 2006, p. 132.

[8] A. Kumar, N. Karnik, and G. Chafle, "Context sensitivity in role-based access control," **Operating systems review**, Vol. 36, No. 3, 2002, pp. 53–66.

[9] P. Eugster, P. Felber, R. Guerraoui, and A. Kermarrec, "The many faces of publish/subscribe," **ACM computing Surveys**, Vol. 35, No. 2, 2003, pp. 114–131.

[10] Mobilegov, "Mobilegov Access Control®" See related information on http://www.mobilegov.com/, 2009.

[11] Y. Kim, C. Mon, D. Jeong, J. Lee, C. Song, and D. Baik, "Context-aware access control mechanism for ubiquitous applications," **Lecture Notes in Computer Science (LNCS)**, Vol. 3528, 2005, pp. 236–242.

[12] R. Thomas and R. Sandhu, "Models, protocols, and architectures for secure pervasive computing: Challenges and research directions," in **Second IEEE Annual Conference on Pervasive Computing and Communications Workshops**, 2004, pp. 164–168.

[13] P. McDaniel, "On context in authorization policy," in **eighth ACM symposium on Access control models and technologies**, 2003, pp. 80–89, ACM.

[14] J. Bacon, K. Moody, and W. Yao, "A model of OASIS rolebased access control and its support for active security," **ACM Transactions on Information and System Security (TISSEC)**, Vol. 5, No. 4, 2002, pp. 492–540.

[15] F. Cuppens and A. Miege, “Modelling contexts in the OrBAC model,” in **19th annual Computer Security Applications Conference**, 2003, pp. 416–425.
[16] A. Westin and O. Ruebhausen, **Privacy and freedom**. Atheneum New York, 1967.
[17] U. Hengartner and P. Steenkiste, “Avoiding privacy violations caused by context-sensitive services,” in **4th IEEE International Conference on Pervasive Computing and Communications, PerCom**, 2006, pp. 222–231.
[18] K. Minami and D. Kotz, “Secure context-sensitive authorization,” **Journal of Pervasive and Mobile Computing**, Vol. 1, 2005, pp. 257–268.
[19] J. Bacon, D. Eyers, J. Singh, and P. Pietzuch, “Access control in publish/subscribe systems,” in **Second International Conference on Distributed Event-Based Systems**, 2008, pp. 23–34.
[20] J. Coutaz, J. L. Crowley, S. Dobson, and D. Garlan, “Context is key,” **Commun. ACM**, Vol. 48, No. 3, 2005, pp. 49–53.
[21] F. Jammes, A. Mensch, and H. Smit, “Service-oriented device communications using the Devices Profile forWeb Services,” in **3rd international workshop on Middleware for pervasive and ad-hoc computing**, 2005, pp. 1–8.
[22] J.-Y. Tigli, S. Lavirotte, G. Rey, V. Hourdin, D. Cheung-Foo-Wo, E. Callegari, and M. Riveill, “WComp Middleware for Ubiquitous Computing: Aspects and Composite Event-based Web Services,” **Annals of Telecommunications (AoT)**, Vol. 64, Apr 2009, pp. 197–214.
[23] C. Perkins, S. Microsyst, and M. Park, “Service Location Protocol for mobile users,” in **ninth IEEE International Symposium on Personal, Indoor and Mobile Radio Communications**, Vol. 1, 1998.
[24] K. Arnold, R. Scheifler, J. Waldo, B. O’Sullivan, and A. Wollrath, **Jini Specification**. Addison-Wesley Longman Publishing Co., Inc. Boston, MA, USA, 1999.
[25] V. Hourdin, J. Tigli, S. Lavirotte, G. Rey, and M. Riveill, “SLCA, composite services for ubiquitous computing,” in **International Conference on Mobile Technology, Applications, and Systems**, 2008, ACM.
[26] G. Kiczales, J. Lamping, A. Mendhekar, C. Maeda, C. Lopes, J. marc Loingtier, and J. Irwin, “Aspect-oriented programming,” in **ECOOP**, 1997, SpringerVerlag.

**Jean-Yves Tigli** got his PhD degree in computer science from the University of Nice Sophia Antipolis, in 1996, on software engineering for intelligent robotics systems. He participated in various European projects between 1998 and 2002 (in ESPRIT and MAST European research programs). He’s Associate Professor in Computer Science at the Engineering School of Technology of the University of Nice – Sophia Antipolis, France. He’s currently managing and leading a project called “Continuum” supported by the French national research agency (ANR) to address the challenge of service continuity in dynamic pervasive environments involving various French universities and international companies.

**Stéphane Lavirotte** got his PhD degree in computer science from the University of Nice – Sophia Antipolis and INRIA, in 2000, on software for document Analysis and Recognition. He participated in various European projects between 1997 and 2004 (in ESPRIT, IST European research programs).
He is Associate Professor in Computer Science at the IUFM of the University of Nice – Sophia Antipolis, France.

**Gaëtan Rey** got his PhD degree in computer science from the University of Joseph Fourrier at Grenoble, in 2005, on context-aware computing. During 2005-2006, he spent one year in the System Research Group of the University College of Dublin,UK. He’s Associate Professor in Computer Science in the Institute of Technology of the University of Nice – Sophia Antipolis, France.

**Vincent Hourdin** is preparing his PhD thesis on context-based security in SOA for pervasive computing at the University of Nice – Sophia Antipolis, supervised by J.-Y. Tigli and Michel Riveill. He’s also software research engineer for MobileGov, an IT security software editor in Sophia Antipolis, France.

**Michel Riveill** got his PhD degree in computer science from the National Polytechnic Institute of Grenoble, in 1987, on distributed software. He obtained “Habilitation à Diriger les Recherches” in 1993, from the same institute. He’s full Professor in Computer Science at the Engineering School of Technology of the University of Nice – Sophia Antipolis, France. He’s leading the computer science department of the engineering school and the software engineering department of the computer science laboratory of the University of Nice - Sophia Antipolis and CNRS.